# Evolution of the Solar Wind Speed with Heliocentric Distance and Solar Cycle. Surprises from Ulysses and Unexpectedness from Observations of the Solar Corona

O. V. Khabarova[a, *], V. N. Obridko[a], R. A. Kislov[a, b], H. V. Malova[b, c], A. Bemporad[d], L. M. Zelenyi[b], V. D. Kuznetsov[a], and A. F. Kharshiladze[a]

[a] Pushkov Institute of Terrestrial Magnetism, Ionosphere and Radio Wave Propagation, Russian Academy of Sciences, Troitsk, Moscow, 108840 Russia

[b] Space Research Institute, Russian Academy of Sciences, Moscow, 117997 Russia

[c] Scobeltsyn Nuclear Physics Institute, Moscow State University, Moscow, 119234 Russia

[d] Istituto Nazionale di Astrofisica (INAF), Osservatorio Astrofisico di Torino, Torino, 10025 Italy

*e-mail: habarova@izmiran.ru

Received April 10, 2018

**Abstract**—An extensive analysis of *Ulysses* observations of the solar wind speed *V* from 1990 to 2008 is undertaken. It is shown that the evolution of *V* with heliocentric distance *r* depends substantially on both the heliolatitude and the solar activity cycle. Deviations from the predicted Parker's profile of $V(r)$ are so considerable that cannot be explained by a scarcity of measurements or other technical effects. In particular, the expected smooth growth of the solar wind speed with *r* is typical only for the solar activity maximum and for low heliolatitudes (lower than $\pm 40°$), while at high latitudes, there are two $V(r)$ branches: growing and falling. In the solar activity maximum, *V* increases toward the solar pole in the North hemisphere only; however, in the South hemisphere, it decreases with heliolatitude. In the minimum of solar activity, the profile of $V(r)$ at low heliolatitudes has a local minimum between 2 and 5 AU. This result is confirmed by the corresponding data from other spacecraft (*Voyager 1* and *Pioneer 10*). Unexpected spatial variations in *V* at low heliolatitudes can be explained by the impact of coronal hole flows on the $V(r)$ profile since the flows incline to the ecliptic plane. To reproduce the impact of spatial variations of *V* in the polar corona on the behavior of *V* at low heliolatitudes, a stationary one-fluid ideal MHD-model is developed with account of recent results on imagery of the solar wind speed in the corona up to 5.5 solar radii obtained on the basis of combined observations from *SOHO/UVCS*, *LASCO*, and Mauna Loa.

## 1. INTRODUCTION

The solar wind speed is a key parameter that determines the dynamics of the heliosphere, namely, the distribution of flows, energy transfer, and the interaction of flows and their evolution with distance. In theoretical works, beginning with the works by Parker [1], the outflow speed of the solar wind is assumed to be radial at all heliolatitudes in the corona and the evolution of the averaged speed profile with distance from the Sun is supposed to be determined by the temperature of the solar corona (see Fig. 1a). In Parker's model, the radial component of the speed grows rapidly and comes to a plateau before it reaches the Earth's orbit. Indeed, the solar wind speed is primarily radial, and its predicted values at 1 AU, on average, match observations, as one can find from Fig. 1a. Even under the consideration of possible kinetic effects that might affect the global behavior of the speed during the solar wind propagation, the plateau in the solar wind speed profile does not disappear [2] and is considered to be a "calling card" of this parameter (see Fig. 1b). As a result of the assumption on a weak dependence of the speed on the heliocentric distance up to the outer heliosphere, the solar wind speed is considered either as a constant or having a profile suggested by Parker in [1]. Alternatively, its averaged observed values are used for setting boundary conditions or for the validation of models (for example, in studies of the evolution of turbulence developing in the solar wind [3]).

A reconstruction of the speed profile in the corona up to $6R_\odot$ (where $R_\odot$ is the radius of the Sun) is possible owing to the combined analysis of coronagraphic measurements in visible light and ultraviolet [4]. Beyond the visibility limits of coronagraphs and up to the areas where in situ measurements begin, the values of the solar wind speed $V$ can be reconstructed through the interplanetary scintillation data [5]. The same method can be used for estimations of $V$ in the heliosphere at different distances [6]. Under rough averaging, observed values of the solar wind speed correspond to modeling [5, 6].

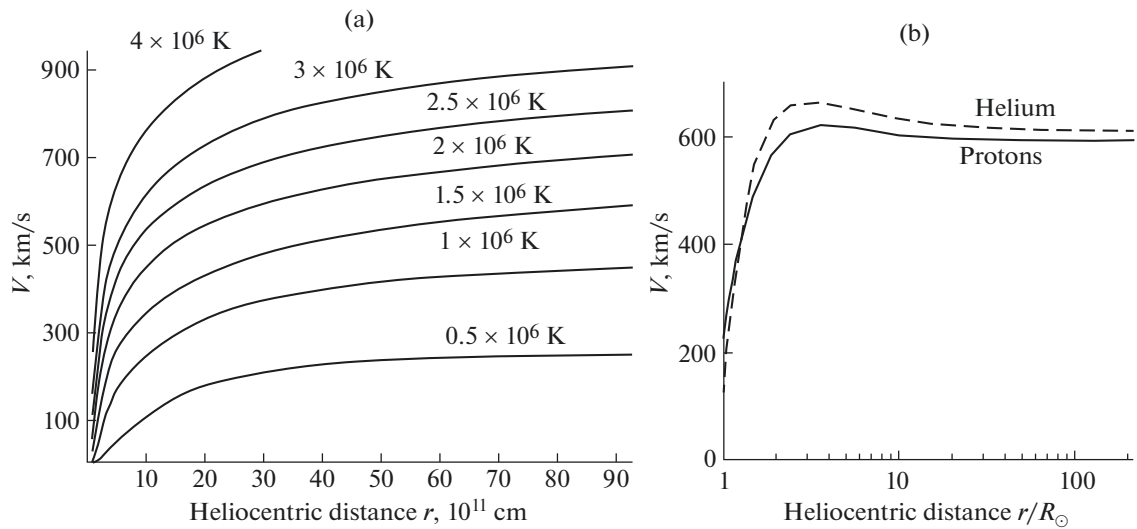

**Fig. 1.** Theoretical views on the radial dependence of the solar wind speed: (a) solar wind speed for different temperatures of the solar corona according to Parker (adapted from [1]), and (b) velocity of alpha particles and protons under taking into account kinetic effects in the solar wind (adapted from [2]).

An interest of specialists in observations of the solar wind speed in the heliosphere considerably increased after the launch of the *Ulysses* spacecraft. *Ulysses* had the orbit practically perpendicular to the ecliptic plane, which allowed observing processes in the solar wind at high heliolatitudes. The first articles on the analysis of the dependence of $V$ on the heliolatitude showed a considerable difference between the solar wind characteristics over the poles and in the ecliptic plane. *Ulysses*, having made the first flyby in the minimum of solar activity, observed a well-pronounced transition from the slow solar wind to the high-speed solar wind related to streams from regions with an open configuration of the magnetic field, in particular, coronal holes [7]. The heliolatitude diagram of $V$ based on the *Ulysses* measurements is characterized by a crown-like increase toward poles in the both hemispheres: the fast solar wind was observed closer to the poles, and the slow wind was detected closer to the equator [7]. The diagram is still the most known profile of the solar wind speed, despite the subsequent publications of the *Ulysses* team showing that this profile changes during the solar cycle considerably. In solar maximum, the difference in the behavior of the solar wind speed at high and low heliolatitudes practically disappears [8].

The latter effect was studied in [9] simultaneously with the analysis of spatial changes in the interplanetary magnetic field (IMF). It was shown that two maxima of the $V$ distribution occur due to the nontrivial dependence of the speed on heliolatitude and solar cycle phase (see Fig. 2). The low-speed maximum of the histogram of distribution shown in Fig. 2a is formed by $V$ values measured in solar activity minima at low heliolatitudes, which are overlapped with high-latitude measurements in solar maxima. The second, high-speed maximum is formed by $V$ values obtained at high heliolatitudes during the years of minimum of the solar activity.

The dependence of $V$ on heliolatitude shown in Fig. 2b is based on the whole body of available *Ulysses* data. Its shape resembles a bird with claws and open wings. Owing to the parameters of the *Ulysses* orbit and time of measurements, the data for high heliolatitudes and the solar activity minima dominate. If *Ulysses* did not exhaust its resources and continued measurements for one more period of solar activity maximum or spent more time at low latitudes, the "claws" in Fig. 2b would be as pronounced as "wings." As a result, the picture would be symmetric, resembling a butterfly or two connected horseshoes. A very similar profile is seen in interplanetary scintillation reconstructions of the dependence of the solar wind speed on heliolatitude (see Fig. 6 in [6]). Unfortunately, the accuracy of this alternative method is low. Furthermore, the scheme of the speed reconstruction is based on two strong assumptions: the solar wind speed is independent of distance and strictly radial. These assumptions can bring a systematic error to the reconstructed two-dimensional picture of the solar wind speed as there is evidence for nonradiality of $V$ at high latitudes even near the solar surface.

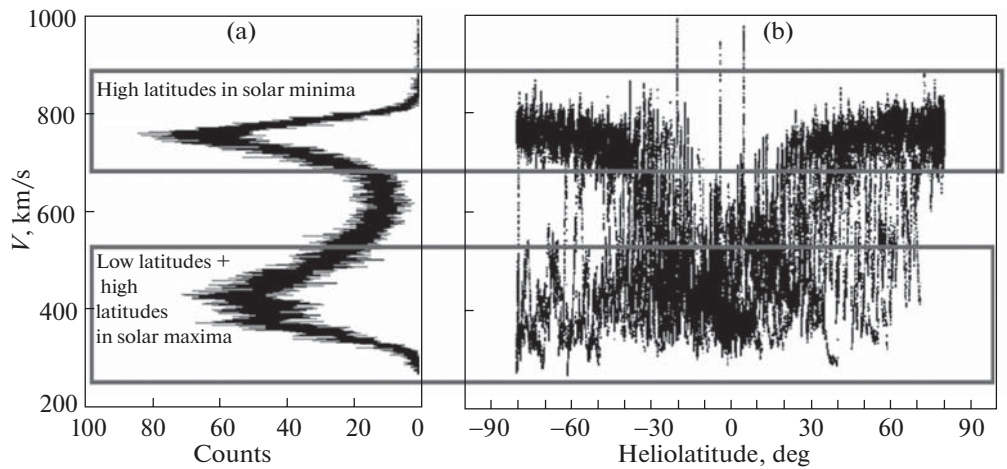

**Fig. 2.** Solar wind speed as observed by *Ulysses* for the entire period of observations (25.10.1990−30.09.2009): (a) histogram (turned by 90°) illustrating the contribution of *V* values obtained at each heliolatitude and (b) latitudinal dependence of *V* [9].

In [10], for instance, an analysis of divergence of the magnetic field lines in the polar corona was performed for solar activity minimum. Polar coronal jets were used as tracers of magnetic field lines. Significant deviations of the magnetic field in the polar corona from the radial direction were found even at distances of one to two $R_\odot$. Taking into account the IMF frozen into the solar corona plasma, it is natural to assume some deviations of the solar wind outflow from the radial direction at the source surface, at least in polar regions of the Sun. Indeed, the idea that flows from coronal holes can reach low heliolatitudes appeared 45 years ago [11], and the basis for this supposition was observations of streams from polar coronal holes at the Earth's orbit in solar minimum. This fact reflects a complex character of the dependence of *V* on distance and heliolatitude, which is almost not studied.

It should be noted that the number of works devoted to observations of the evolution of *V* with distance in the heliosphere is very small. In most cases, the behavior of *V* is analyzed through the calculation of the dispersion of values obtained from measurements of a few spacecraft with an elongated orbit, and a statistical analysis of *V* at several selected distances is sometimes performed [6−8, 12]. However, both approaches are inapplicable to the comparison with Parker-like models that provide results in a form of the curves shown in Fig. 1.

In this work, we fill the existing gap and analyze the dependence of the solar wind speed on distance and heliolatitude, taking into account the phase of the solar cycle. An analysis performed in this study has shown the existence of a local minimum of the solar wind speed at distances of several AU in the minimum of solar activity. In order to check the obtained result, data from alternative spacecraft have been examined, and the examination confirmed the found effect. A two-dimensional MHD model has been built for its explanation. We show that the solar wind speed profile at low heliolatitudes significantly depends on spatial changes of the speed in the entire corona, including the polar regions of the Sun.

## 2. RESULTS OF OBSERVATIONS

### 2.1. Data and Method

As noted above, *Ulysses* is the only spacecraft in the history of space explorations that studied the heliosphere not near the ecliptic plane as other missions, but practically perpendicular to it. The parameters of its orbit and key solar wind characteristics observed during the *Ulysses* flybys of the Sun can be found in Fig. 3 from [9]. To study temporal and spatial variations of *V* in key phases of the solar cycle, we selected years in the vicinity of the solar activity minimum (1995−1997, 2007−2009) and maximum (1990−1991, 2000−2002).

Performing an analysis on the radial dependence of the speed, one should take into account the existence of a zone of the increased solar activity within the band of ±40° around the solar magnetic equator, known since the time of Maunder [13−16]. The occurrence of the region is reflected in the behavior of the IMF and the solar wind speed [6, 9, 17]. Therefore, it is reasonable to divide the heliosphere into high-latitude and low-latitude at the border of 40° heliographic latitude. Afterwards, the radial dependence of the solar wind speed in a certain phase of the solar cycle can be found and analyzed in each zone. We additionally study the dependence of the solar wind speed on heliolatitude, $V(\theta)$. Since we are interested in a general view of the dependences, we employ the daily *Ulysses* data, according to which it is possible to calculate necessary approximations. To check the effects found from the analysis of *Ulysses* measurements, we employ data from other spacecraft and obtain an analogue of the curve $V(r)$ shown in Fig. 1, averaging the hourly $V$ data per 1 AU beyond the Earth's orbit in the way introduced in [17]. Under such an approach, all nonstationary effects (from short-term turbulent disturbances to the impact of the interplanetary coronal mass ejections) become smoothed and insignificant.

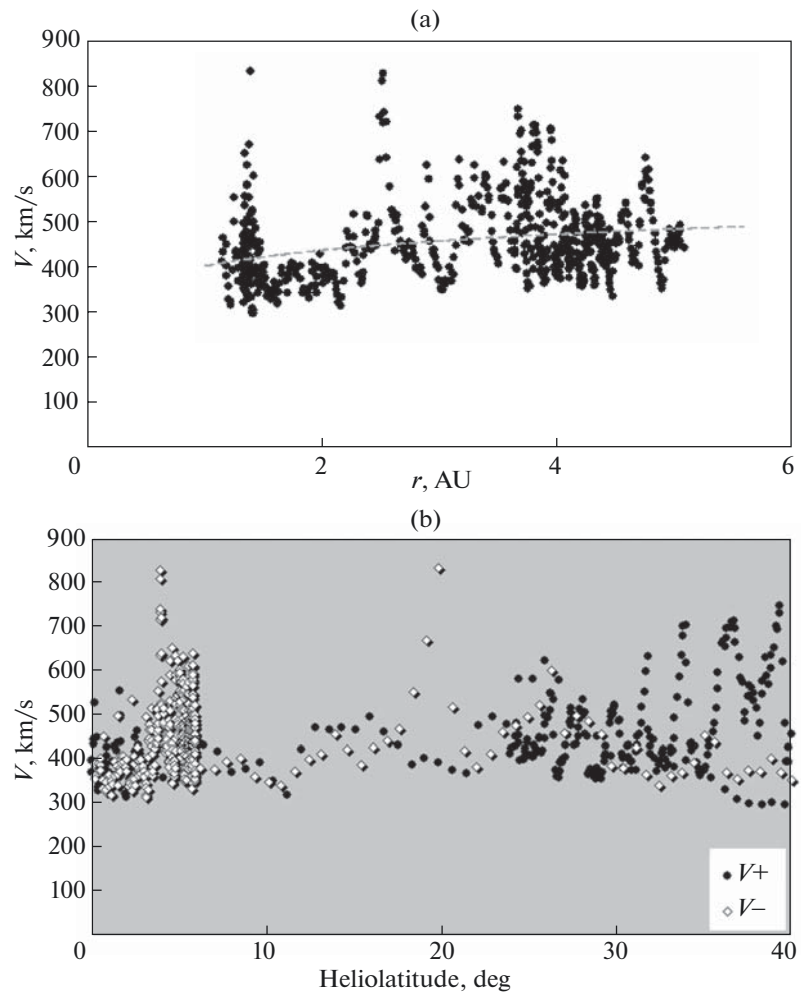

**Fig. 3.** Solar wind speed observed by *Ulysses* in years of maximum of the solar activity at low heliolatitudes within the belt of ±40° heliographic latitudes with respect to the helioequator: (a) radial dependence $V(r)$ and (b) dependence on absolute value of the heliolatitude $V(\theta)$. Here and below, the data for the northern and southern hemispheres are indicated by black dots ($V+$) and white diamonds ($V-$), respectively.

### 2.2. Low Heliolatitudes

*Ulysses* was launched in a year of solar maximum. It performed a passage along the ecliptic plane toward the Jupiter, and further, at the approach to Jupiter to 6.3 Jovian radii in the beginning of February 1992, its orbit was lifted as a result of the gravitational maneuver. Due to this, a full coverage of the profile of $V$ from the Earth to several AU at low heliolatitudes was pos-sible only in the solar maximum of 1990−1991. During the following maximum of solar activity (2000−2002), *Ulysses* was at low heliolatitudes only while coming closer to the aphelion and the perihelion. The spatial profiles $V(r)$ and $V(\theta)$ obtained in maxima of solar activity within the area of ±40° above/below the helioequator from all available *Ulysses* data are shown in Figs. 3a and 3b, respectively.

The $V(r)$ profile meets expectations of a smooth growth of the solar wind speed and reaching a kind of plateau with distance, as seen in Fig. 3a when comparing with Fig. 1. Hereinafter, the dotted line is an approximation by a third-order polynomial. The high density of dots in the vicinity of low heliolatitudes in Fig. 3b is a consequence of a prolonged stay of the *Ulysses* spacecraft in the ecliptic plane while approaching Jupiter, up to the gravitational maneuver. There is no any considerable difference between the behavior of the $V(r)$ profiles in the northern and southern low-latitude heliosphere (positive and negative heliolatitudes, respectively) in the years of solar maximum at least up to 35° of heliolatitude. Also, there is no obvious dependence of $V$ on heliolatitude at low heliolatitudes, despite the expressed variability of $V(\theta)$. Therefore, in the maximum of solar activity, the solar wind speed behaves at low heliolatitudes according to theoretical predictions.

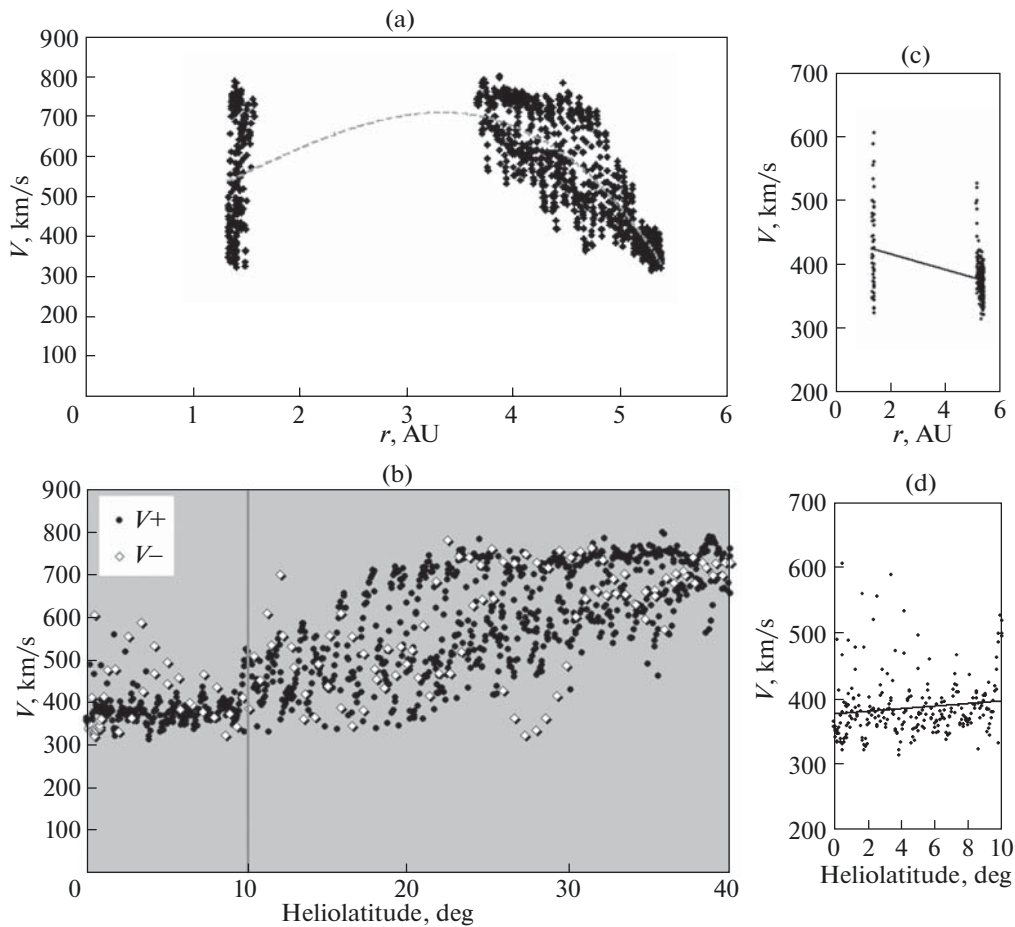

**Fig. 4.** Solar wind speed observed by *Ulysses* in years of minimum of the solar activity at low heliolatitudes within the belt of ±40° with respect to the helioequator: (a) radial dependence $V(r)$ and (b) dependence on absolute value of the heliolatitude $V(\theta)$. Right panels (c) and (d) show the data extraction from (a) and (b) within the ±10° band around the helioequator.

Since solar maximum is treated as the most problematic period for comparisons of a theory with observations, there is apparently every reason to see a fuller correspondence of the spatial picture of the speed in the heliosphere to theoretical expectations in the minimum of solar activity. However, in solar minimum, the situation cardinally changes toward a striking disagreement between observations and expectations, as seen from Fig. 4. The corresponding profiles of $V(r)$ and $V(\theta)$ can be found in Figs. 4a and 4b, respectively. Despite the data gap at some distances owing to specific features of the spacecraft orbit, the first peculiar-ity that attracts attention in Fig. 4a is the falling profile of the radial dependence of $V$ in the solar activity minimum.

The $V(\theta)$ profile also changes completely (see Fig. 4b). There is a zonality even within the selected ±40° band of the low-latitude solar wind, i.e., there is a clear difference between the behavior of the solar wind speed near the ecliptic plane and above (±10° with respect to the heliographic equator). Within the ±10° belt, the speed behaves similar to Fig. 3b; however, there is a steady trend to a smooth increase in the solar wind speed with heliolatitude outside the belt.

To avoid the impact of the revealed $V(\theta)$ dependence on $V(r)$ profile and emphasize the found trend of $V$ to decrease with distance in the solar minimum, it is reasonable to restrict the same analysis by the ±10° belt, i.e., to use data around the *Ulysses* perihelion and aphelion. Results of this selection and the linear approximation are shown in Figs. 4c and 4d, to the right of the corresponding panels of Figs. 4a and 4b. It is easy to find that $V$ falls from the average value of about 550 km/s at 1.34 AU to 350 km/s at 5.37 AU even under the reducing number of points, i.e., the trend to the decrease instead of the expected growth is obvious.

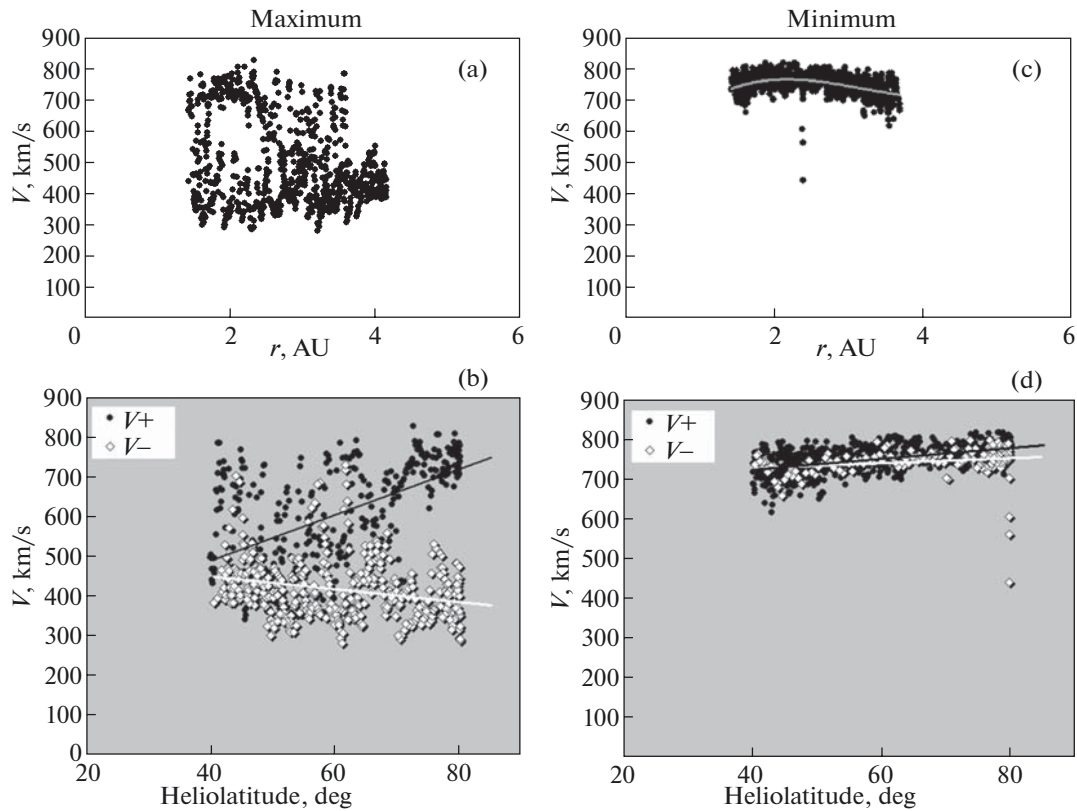

**Fig. 5.** Solar wind speed observed by *Ulysses* in years of maximum (a, b) and minimum (c, d) of the solar activity at high heliolatitudes above 40°: (a) radial dependence $V(r)$ and (b) dependence on absolute value of the heliolatitude $V(\theta)$.

The occurrence of the zonality can be explained easily with the existence of the well-developed streamer belt forming a separated area near the heliospheric current sheet in the solar minimum. However, both found phenomena, namely, a smooth growth of the $V(\theta)$ profile outside of the $\pm 10°$ area (Fig. 4b) and the decrease in $V(r)$ beyond 1 AU near the ecliptic plane (Figs. 4a, 4c), need a careful studying.

In solar minimum, long-lived high-latitude coronal holes occur, and the corresponding flows dominate in the heliosphere. Therefore, a considerable impact of factors related to processes in the high-latitude heliosphere on the low-latitude solar wind become obvious. To understand it, we will continue with the analysis of *Ulysses* data at high heliolatitudes.

### 2.3. High Heliolatitudes

High heliographic latitudes are characterized by a complicated radial dependence of $V$ in the solar maxi-mum (see Fig. 5a). The solar wind speed profile possesses two branches here, one of which quickly falls with distance, while the other grows like the expected $V(r)$ profile (see the corresponding plot for the low heliolatitudes in Fig. 3a for comparison).

The existence of the two branches is understandable at the accounting of the striking difference between the behavior of $V(\theta)$ in the northern and southern heliospheres. Figure 5b shows that, while approaching the pole, $V$ grows in the northern hemisphere (corresponding to positive values of the heliolatitude) and falls in the southern hemisphere. Thus, the difference between the two branches is so considerable that cannot be accidental.

This effect can be explained by the inclination of the magnetic axis of the solar dipole, by the impact of the quadrupole magnetic field in solar maximum, or by the existence of the dominating hemisphere. Independently of its nature, the occurrence of the steady falling branch seen in Fig. 5a is a signature of a far more difficult dependence of the solar wind speed on distance and heliolatitude than it was supposed before.

In the solar minimum at high heliolatitudes, only one more or less ordered $V(r)$ branch with a maximum and falling edges remains (see Fig. 5c). The $V(\theta)$ pro-file becomes simpler as well (Fig. 5d). The way of the $V(\theta)$ profile growth differs in the southern and northern hemisphere while approaching polar regions. The falling branch in the southern heliosphere found in Fig. 5b turns here into the growing.

It is possible to interpret the observed profile of the solar wind speed at high latitudes by the occurrence of polar coronal holes in the solar minimum and the impact of the dependence of $V$ on the distance to the coronal hole center. It is known that the difference between $V$ values at the coronal hole border and its center is ~50–100 km/s for the solar minimum [18]. It coincides by order of magnitude with changes in the $V$ profile seen in Fig. 5c and 5d. Meanwhile, the assumption on the contribution of coronal hole flows to the observed $V(r)$ profile requires a further examination, as polar coronal holes were not well-developed during some periods of *Ulysses* observations in solar minima.

## 3. CHECK OF THE EFFECT OF A LOCAL SOLAR WIND SPEED DECREASE WITH DATA FROM SEVERAL IN-ECLIPTIC SPACECRAFT

To check whether the effect of the decrease in the solar wind speed observed by *Ulysses* beyond the Earth's orbit in the minimum of solar activity is random or not, we will employ a well-proven technique used for the analysis of spatial changes in the IMF [17]. Due to the practically simultaneous occurrence of some spacecrafts near the ecliptic plane at different distances from the Sun in the "golden age" of space exploration, it is possible to obtain data on the characteristic values of the solar wind parameters over a considerable distance. We use a database of six spacecraft, namely, *Helios 1*, *Helios 2*, *Pioneer Venus Orbiter* (*PVO*), *IMP8*, and *Voyager 1* from the list [17], adding *Pioneer 10*. The first four spacecraft provided data from 0.29 to 1 AU from 1976 to 1979 (the minimum and the growing phase of the solar cycle), and *Pioneer 10* was launched in 1972, which corresponded to the decreasing phase of the solar activity. Therefore, it is possible to estimate the behavior of the solar wind speed in the vicinity of a solar minimum. Hourly data have been averaged for 1 AU beyond the Earth's orbit and for 0.2 AU closer to the Sun (see Table 1 for details).

To compensate a possible effect of the *Voyager 1* and *Pioneer 10* movement with regard to the solar wind, the values of the radial component of the spacecraft speed averaged over 1 AU were added to the averaged $V$ values at the corresponding distances. There was no such correction made for spacecraft with the orbit inside 1 AU, as the averaging made for *Helios 1* and *Helios 2* compensates the impact of the spacecraft back-and-forth movement with respect to the radial direction and *PVO* and *IMP8* are fixed at one point against the radial propagation of the solar wind at the scales of averaging. Our calculations show that the correction for *Voyager 1* and *Pioneer 10* is insignificant. The corresponding values vary from 14 to 8 km/s. Because of this, the movement of spacecraft with elongated orbits is usually ignored in the analysis of the low-resolution $V$ data [19]. However, we considered the necessity of taking into account all possible technical effects to show the occurrence of a deep decrease in the solar wind speed at several AU.

**Table 1.** Number of hourly measurements of $V$ contributing to the averaged values calculated for the corresponding spatial intervals shown in Fig. 6

| AU | Pioneer 10 | Voyager 1 | Helios 1 | Helios 2 | IMP8 | PVO |
|---|---|---|---|---|---|---|
| 0.33 | | | | 4155 | | |
| 0.34 | | | 3773 | | | |
| 0.5 | | | 5297 | 4136 | | |
| 0.7 | | | 6368 | 5384 | | 6973 |
| 0.9 | | | 12043 | 12806 | | |
| 1 | | | | | 15199 | |
| 1.5 | 1679 | 2100 | | | | |
| 2.5 | 2334 | 845 | | | | |
| 3.5 | 2588 | 1324 | | | | |
| 4.5 | 3938 | 2241 | | | | |
| 5.5 | 4754 | 3641 | | | | |

Figure 6 shows radial profiles of $V$ obtained from several spacecraft data (see the list of spacecraft and corresponding designations to the right of Fig. 6). Despite the difference in years of launch and orbits of the spacecraft, the average values of $V$ correspond to each other very well. Even keeping in mind the performed adjustment for the own speed of a spacecraft, one can notice an obvious decrease in the solar wind speed measured by *Voyager 1* and *Pioneer 10* at distances between 1 and 4 AU. Taking into account the number of points used for calculations of averaged $V$ values (see Table 1) and the fact that curves for *Voyager 1* and *Pioneer 10* almost coincide in the region of interest, it becomes clear that the tendency to the solar wind slowdown beyond 1 AU cannot be random.

## 4. MODELING

As follows from Figs. 4 and 6, the radial profile of the solar wind speed is nonmonotonic. Near the solar activity minimum, $V$ has a local minimum beyond 1 AU, in the region of stream–stream interactions. This effect can be a reflection of spatial inhomogeneity of the solar wind outflow speed in the solar corona and show that the solar wind is differently accelerated at different latitudes near the Sun. We will show below that the solar wind speed is a function of the magnetic flux; therefore, it is conserved along IMF lines. In the model, we suggest that magnetic field lines come to low latitudes far from the Sun, becoming closed through the equator. When a spacecraft moves along the ecliptic, magnetic field lines from higher and higher latitudes are crossed further and further from the Sun. As a result, $V$ depends on heliocentric distance as nonmonotonically as it depends on the heliolatitude near the Sun. The model described below illustrates an idea about the projection of the solar wind outflow speed along magnetic field lines from the corona into the heliosphere.

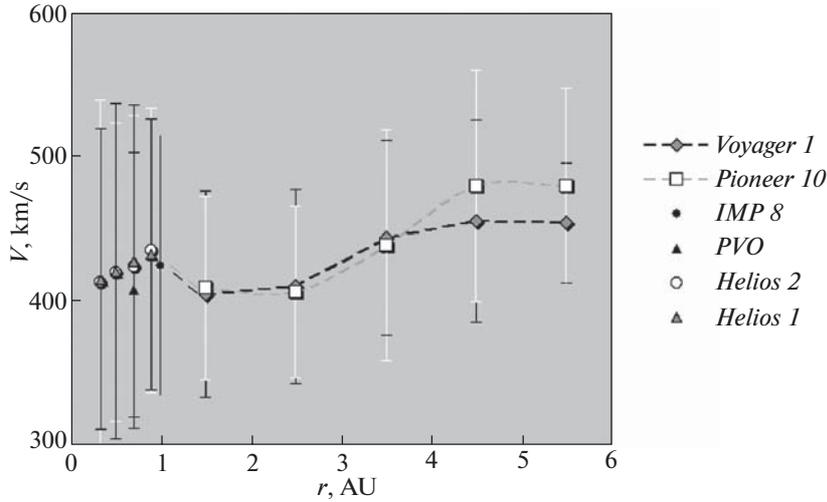

**Fig. 6.** Solar wind speed radial evolution along the ecliptic. Hourly data for the solar minimum and raising phase of the solar cycle from several spacecraft are used (see the corresponding notations to the right of the plot). The vertical lines denote the standard deviation at each point. The local decrease in the solar wind speed observed by *Ulysses* (see Figs. 4a, 4c) is confirmed by measurements from two spacecraft near the ecliptic with a time difference of several years. The number of measurements contributing to each point is shown in Table 1.

### 4.1. Equations

In this section, we will show that the dominating radial component of $V$ is a function of the magnetic flux. The solar wind will be considered within the frame of a stationary one-fluid ideal MHD model. We will choose the spherical coordinate system $(r, \theta, \varphi)$ with the origin in the center of the Sun, where the polar angle $\theta$ is counted from the North Pole. All equations will be presented in the inertial reference system. It is worth noting that we are not going to build a model of the solar wind itself, but we only focus on the qualitative explanation of the effect. Therefore, it is unnecessary to solve a system of MHD equations fully as, e.g., in [20], but it is possible to introduce some simplifications. We will assume that the solar wind is approximately axisymmetric. Effects related to asymmetry can be considered as perturbations of the symmetric plasma balance. Despite being significant during the high solar activity periods, such effects will not be considered in this work.

The full system of equations of stationary one-fluid MHD without heat transfer has the following appearance. The pressure $P$, the plasma temperature $T$, and the number density $n$ are connected by the equation of state,

$$P = nT. \qquad (1)$$

The equations of plasma balance are

$$\rho(\mathbf{v} \cdot \nabla)\mathbf{v} = -\nabla P + \frac{1}{c}[\mathbf{j} \times \mathbf{B}], \qquad (2)$$

$$\nabla \cdot \mathbf{B} = 0, \qquad (3)$$

$$\nabla \times \mathbf{B} = \frac{4\pi}{c}\mathbf{j}, \qquad (4)$$

$$\nabla \times \mathbf{E} = 0 \;\rightarrow\; \mathbf{E} = -\nabla\Psi, \qquad (5)$$

$$\mathbf{E} + \frac{1}{c}[\mathbf{v} \times \mathbf{B}] = 0, \qquad (6)$$

$$\nabla \cdot (\rho\mathbf{v}) = 0. \qquad (7)$$

The designations in Eqs. (1)–(7) are standard: $\mathbf{E}$ is the electric field, $\mathbf{B}$ is the magnetic field, $P$ is the thermal pressure, $T$ is the temperature, $\rho = m_p n$ is the mass density, $m_p$ is the proton mass, $n$ is the plasma number density, $\mathbf{v}$ is the speed, $\mathbf{j}$ is the current density, and $\Psi$ is the electric potential. In addition, the designation for the poloidal part of any vector containing only the radial and tangential $\theta$ components can be used. For example, for the magnetic field it is $\mathbf{B}_p$, while for the speed it is $\mathbf{v}_p$.

We introduce the magnetic flux $\Phi$ [21] as follows:

$$B_r = \frac{1}{r^2 \sin\theta}\frac{\partial\Phi}{\partial\theta}, \qquad (8)$$

$$B_\theta = -\frac{1}{r \sin\theta}\frac{\partial\Phi}{\partial r}. \qquad (9)$$

We will consider the azimuthal projection of "frozen-in" equation (6) taking into account the axial symmetry, $[\mathbf{v} \times \mathbf{B}]_\varphi = 0 \Rightarrow \mathbf{v}_p \parallel \mathbf{B}_p$. Thus, the poloidal magnetic field and the speed are parallel.

Let us consider the longitudinal projection of plasma balance equation (2),

$$\rho B_r \left( v_r \frac{\partial v_r}{\partial r} - \frac{v_\varphi^2 + v_\theta^2}{r} + v_\theta \frac{\partial v_r}{r \partial \theta} \right)$$

$$+ \rho B_\theta \left( v_r \frac{\partial v_\theta}{\partial r} + v_\theta \frac{\partial v_\theta}{r \partial \theta} + v_r \frac{v_\theta}{r} - \frac{v_\varphi^2}{r \tan \theta} \right)$$

$$= -(\mathbf{B} \cdot \nabla)\left( P + \frac{B^2}{8\pi} \right) \qquad (10)$$

$$+ \frac{B_r}{4\pi}\left( B_r \frac{\partial B_r}{\partial r} - \frac{B_\theta^2 + B_\theta^2}{r} + B_\theta \frac{\partial B_r}{r \partial \theta} \right)$$

$$+ \frac{B_\theta}{4\pi}\left( B_r \frac{\partial B_\theta}{\partial r} + B_\theta \frac{\partial B_\theta}{r \partial \theta} + B_r \frac{B_\theta}{r} - \frac{B_\varphi^2}{r \tan \theta} \right).$$

Equation (10) can be substantially simplified if we take into account the orders of the variables in the expression. Observations show that the solar wind is generally radial, i.e., $v_r \gg v_\theta$, $v_\varphi$. As $v_p \parallel B_p$, it follows from the solar wind being directed radially that $B_r \gg B_\theta$ in the used system of coordinates. Therefore, it is possible to reduce Eq. (10) to the following form: $\rho(B,\nabla)v_r^2/2 = -(\mathbf{B} \cdot \nabla)(P + B^2/8\pi) + (4\pi)^{-1}(\mathbf{B} \cdot \nabla)B_r^2/2 - B_r(4\pi r)^{-1}B_\varphi^2$. The obtained expression can be simplified even more, having assumed that the dynamic pressure in the solar wind dominates over the thermal and magnetic pressures, i.e., $\rho v_r^2 \gg P$, $B^2/8\pi$. The last inequality is often used in kinematic models of the solar wind, e.g., in [1]. Finally, we obtain the relationship $\rho B_r \left( v_r \frac{\partial}{\partial r} + v_\theta \frac{\partial}{r\partial\theta} \right) v_r = 0$. Taking into account that $v_p \parallel B_p$, this ratio can be rewritten in the form

$$(\mathbf{B} \cdot \nabla)v_r = 0. \qquad (11)$$

It follows from Eq. (11) that $v_r = U(\Phi)$, where $U(\Phi)$ is any continuously differentiable function of the magnetic flux, which is set by boundary conditions at some distance from the Sun. Thus, we have shown that the radial component of the solar wind speed $v_r$ is unambiguously defined by the magnetic flux. Therefore, the speed is defined by the boundary condition on $U$, and its direction is defined by the magnetic field.

### 4.2. Boundary Conditions

We have shown that, under the simplifications suggested in the model, the radial component of $V$ is a function of the magnetic flux, $U = U(\Phi)$, where

$\Phi = \Phi(r, \theta)$. It is reasonable to set boundary conditions at two surfaces: at the sphere of a finite radius $r_1$ around the Sun and in the equatorial cross section $\theta_1 = \pi/2$. Thus, it is necessary to know the type of the corresponding functions $\Phi_1(\theta) = \Phi(r = r_1, \theta)$ and $\Phi_2(r) = \Phi_0(r, \theta = \theta_1)$. For this purpose, we will consider the spatial distributions of the speed obtained from observations [4] and shown in Fig. 7.

Figure 7a displays the profile of the solar wind in the corona reconstructed from observations using the technique described in [4]. Variations in $V$ vs. polar angle are shown at four selected distances from the Sun. Local decreases in the speed are clearly seen near the equator at 90° and 270°, where the solar wind is less accelerated, as was also shown in [4]. The North Pole corresponds to 0° and 360°, and the South Pole corresponds to 180°. Closer to the poles, the solar wind is already formed between $4.5R_\odot$ and $5.5R_\odot$. One can notice small decreases in the solar wind speed located almost symmetric with regard to the poles. These decreases and their internal structure become especially well pronounced with distance, being completely formed by $5.5R_\odot$. The position of all the peculiarities is stable and can be traced along the radius with a deviation of no more than 1°.

The borders of coronal holes can be identified in Fig. 7a as symmetric drops indicated by vertical lines 1. Small-scale speed decreases (marked by lines 2) in the nearest vicinity of the pole in Fig. 7a are related to the existence of polar conic current sheets observed inside coronal holes in the minimum of solar activity. The combined impact of local and global irregularities of the solar magnetic field on the entire picture of the solar wind streams was suggested in [22]. The existence of separated conic areas at high latitudes of rotating astronomical objects with a strong magnetic field is also known for a long time [23, 24]. The formation of isolated conical magnetic structures inside polar coronal holes on the Sun was predicted in [25], and the structures representing steady conical current sheets were found in the polar heliosphere far from the Sun [26]. These structures are also clearly visible in the picture of reconstructed magnetic field lines in the corona (see Fig. 7b). The description of the technique of obtaining such pictures can be found in [27]. The borders of polar coronal holes are visible not only in plasma parameters, but also in magnetic fields. One can find that the open magnetic field lines adjoining with closed field lines extend and incline to the equator yet in the corona.

The function $U$ is defined by the magnetic flux $\Phi(r, \theta)$. At the sphere of radius $r_1 = 4.5R_\odot$, corresponding to the "beginning" of magnetic field lines, it is defined by the stream function $\Phi_1(\theta)$. At $\theta = \pi/2$, i.e., at the equator, where the "ends" of magnetic field lines occur, the magnetic flux is $\Phi_2(r)$. The substitution of $\Phi_2$ into $U(\Phi)$ allows one to determine the type of the $U(r)$ dependence. We will set a form of the function $\Phi_1(\theta)$ under the assumption that the magnetic field of the Sun is dipole,

$$\Phi_1(\theta) = \Phi_0 \sin^2 \theta. \quad (12)$$

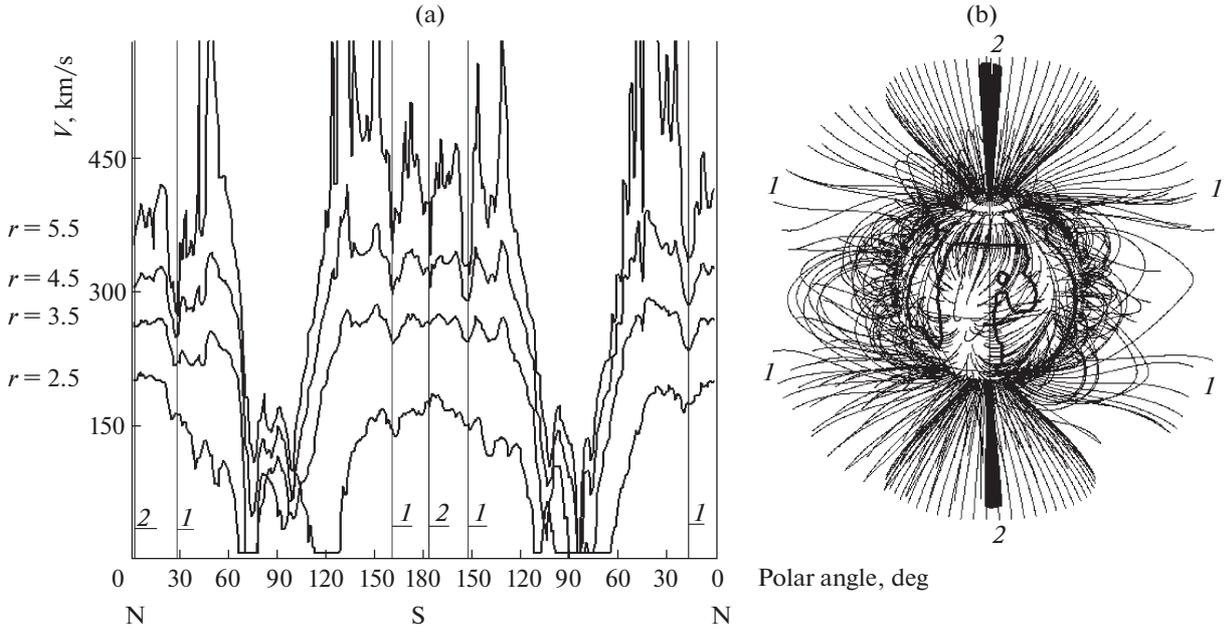

**Fig. 7.** Profiles of the speed and magnetic field in the corona. (a) Example of the dependence of the solar wind speed on the polar angle at different distances (in solar radii) observed on June 9, 1997. Line *1* is the border of the coronal hole, and line *2* indicates the conic polar current sheet fully formed at $r > 4.5 R_\odot$. (b) Extrapolation of magnetic field lines from the photosphere to the source surface at $3.25 R_\odot$ for the same period. The numerals mark the same regions as in panel (a). The magnetic field lines located close to the coronal hole boundary incline to the lower heliolatitudes. Despite dynamicity of the solar magnetic field, peculiarities *1* and *2* can be clearly distinguished through the entire solar minimum.

Here, $\Phi_0$ is the full magnetic flux through the northern hemisphere of the Sun at $r = 4.5 R_\odot$, normalized to $2\pi$. The dependence of the radial IMF component on the distance at low latitudes is found to be $B_r \sim r^{-5/3}$ according to observations from several spacecraft [9, 17]. To choose the positions of "ends" of magnetic field lines at low heliolatitudes, we put $B_\theta \sim r^{-8/3}$, taking into account Eq. (8). The latter qualitatively corresponds to information on the spatial distribution of the magnetic field in similar magnetic structures in magnetospheres of planets, e.g., Jupiter [23, 24]. Finally, we obtain the following expression:

$$\Phi_2(r) = \Phi_0 (r/R_\odot)^{-2/3}. \quad (13)$$

The function $\Phi_0$ entering into Eqs. (12) and (13) can be excluded by the normalization of both functions. It should be noted that the real magnetic field of the Sun can have a significant quadrupole component [22]. However, we do not take it into account for simplicity, as the quadrupole component rapidly decreases with distance and has no significant impact on $\Phi_2(r)$. The given boundary conditions allow calculating the solar wind speed in the region of interest.

### 4.3. Solutions

Figure 8 shows the chosen boundary conditions (Fig. 8a) and the obtained solutions for $V(r)$ far from the Sun (Fig. 8b) for the projection of the boundary conditions onto the near-equator area of the heliosphere. The boundary conditions shown in Fig. 8a correspond to Fig. 7a at $r = 4.5 R_\odot$. Figure 8b reveals a strongly nonmonotonic radial evolution of the solar wind speed, showing a big difference between the northern and southern heliosphere. One can find that two branches of $V(r)$ anticorrelate in different hemispheres. The solar wind speed is higher in the southern hemisphere approximately up to 2.1 AU; however, the solar wind further propagates faster in the northern hemisphere up to 6.5 AU, and then the situation repeats. In the northern heliosphere, the extremum occurs approximately at 1.5 AU, while in the southern heliosphere, the extremum is reached at 2.5 AU. There is an additional peak near the Earth's orbit in the southern low-latitude heliosphere.

Lines 1N and 1S in Fig. 8b show the features caused by the existence of the border of coronal hole in the corresponding hemispheres. It should be noted that conical current sheets *2* in Fig. 7 are projected onto the edge of the heliosphere and are not responsible for the solar wind speed variations at low heliolatitudes. In reality, one can expect that they do not reach low latitudes at all or contribute to the spatial picture of the solar wind speed at very far heliocentric distances, out of the frame of Fig. 8b. Therefore, the picture of spatial variations of the solar wind speed observed at the lower heliolatitudes depends not only on the formation of the speed profile in the corona near the solar equator but also on the speed spatial peculiarities in the polar corona.

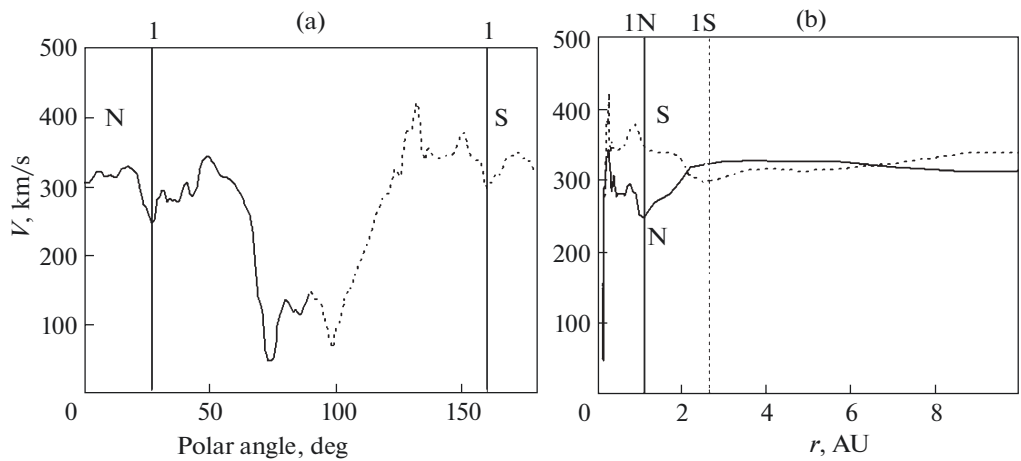

**Fig. 8.** (a) Spatial variations of the solar wind outflow speed in the corona (at $4.5R_\odot$, where boundary conditions are set) and (b) corresponding solutions for $V$ near the ecliptic. Curves N and S correspond to the northern hemisphere (the solid line) and southern hemisphere (the dotted lines), respectively. Line $1$ indicates the boundary of a coronal hole, similar to Fig. 7.

## 5. DISCUSSION AND CONCLUSIONS

An understanding of specific features of spatial variations in the solar wind speed in the heliosphere is an extremely important task, as $V$ is one of the key parameters used in all models that describe solar wind processes. For example, the solar wind speed plays a crucial role when studying the time evolution of turbulence in the heliosphere [3, 28]. Meanwhile, the views on the evolution of the solar wind speed with distance and solar activity cycle still remain quite primitive. It is usually supposed in models that (a) $V$ rapidly grows up to the Earth's orbit and further remains almost unchanged; (b) the solar wind speed is radial; and (c) the high-latitude solar wind is fast, while the low-latitude solar wind is slow. Thus, it is commonly accepted to separate the high-speed and low-speed solar wind simply by values of $V$ (usually, the threshold is selected to be 450 km/s), without understanding the origin of certain solar wind streams and accepting the postulate that the properties of the solar wind depend on the value of $V$ only.

This approach works rather well near the ecliptic plane, where such a criterion is reasonable and based on the knowledge of the properties of solar wind streams observed mainly at 1 AU [29]. However, when one analyzes data of spacecraft measuring the solar wind parameters far from Earth or far from the ecliptic plane, it can lead to an incorrect interpretation of the obtained results.

In the *Ulysses* case, the separation into the fast and slow solar wind simply by the speed value is meaningless, as the speed depends on both heliolatitude and solar cycle. As a result, these dependences contribute to the correlation between $V$ and any other parameter, which changes the result cardinally. Imagine that one wants to find, for instance, whether the magnetic flux depends on the solar wind speed or not. Treating the two humps of the $V$ distribution in Fig. 2a as the "slow" and "fast" solar winds, one actually obtains the characteristics of the magnetic flux in different phases of the solar cycle and at different latitudes instead of finding the dependence of the magnetic flux on $V$. The phase–latitude dependences form a heliolatitude chart of the solar wind, as is shown in Fig. 2b (see examples in [9]). The dependence of the magnetic flux on the phase of the solar cycle is a widely known fact [15, 16]. A weak dependence of the magnetic flux on heliolatitude has also been found recently in [9]. However, it is difficult to extract these two effects from the *Ulysses* speed data without knowing anything about the $V(r)$ and $V(\theta)$ profiles, as well as about the dependence of $V$ on the solar activity cycle. As a result, it is easy to make an incorrect conclusion on the existence of the dependence of the magnetic flux on the solar wind speed.

In this work, we have filled the gap in studying the solar wind properties by analyzing the solar wind speed evolution with distance, latitude, and phase of the solar cycle based on the *Ulysses* measurements. Considerable deviations of observations from theoretical predictions at several AU were found. *Voyager 1* and *Pioneer 10* data were used to check a local deceleration of the solar wind beyond the Earth's orbit as revealed by *Ulysses*. The examination confirmed a nonrandomness of the effect. Unfortunately, as noted above, previous works on this subject were insufficient in number and did not give an opportunity to estimate if there are any deviations from the picture of spatial evolution of the solar wind speed suggested by Parker (Fig. 1a). Previous attempts to connect variations in $V$ observed by several spacecraft with the phase of the solar cycle or heliocentric distance [12, 19, 30, 31] did not show the radial dependence of $V$ in the classic Parker's form. A prior analysis was carried out in a form of histograms or direct comparisons of data rows sometimes smoothed but abounding with numerous outlying points caused by nonstationary processes, etc. Furthermore, an attempt to pay attention to a strange decrease in the solar wind speed in the course of solar wind propagation to the outer heliosphere was made only in [31]. The solar wind speed variations closer to the Sun than 5 AU were not discussed and considered completely meeting expectations. The authors of [31] explained the solar wind deceleration at large heliocentric distances by the impact of ions from the interstellar medium (so-called pickup ions). It was suggested that the 7% decrease in the solar wind speed beyond 10 AU is related to a decrease in the density because of the replacement of thermal solar wind ions by pickup ions. It was claimed that it is, in fact, a technical effect, because a spacecraft counts only thermal ions which become not recognized due to the charge exchange and replacement by hot pickup ions with distance. It was supposed that pickup ions make 6–8% of all solar wind ions beyond 5 AU. According to the logic of [31], the 8% decrease in the density causes a decrease in the observed solar wind flux and, consequently, the speed. Not discussing this interpretation here, we note that the "replacement" effect is insignificant in the inner heliosphere and cannot be employed to explain the observed decrease in the $V(r)$ profile at several AU.

It is also impossible to explain Fig. 6 by the heliolatitude/longitudinal dependence related to short-term solar activity variations, i.e., temporal and random $V$ changes on the Sun at the time of observations, as was done, e.g., in [32]. First, the average is made over a great body of measurements covering large temporal and spatial intervals (see Table 1 for details). Second, the nonrandomness of the decrease in the solar wind speed beyond the Earth's orbit is confirmed by measurements of two spacecraft with different characteristics, different orbits, and with a five-year difference in launches. The spacecraft provide almost identical average values of the speed that lie unambiguously below the curve that meets theoretical expectations on a smooth growth of $V$ and plateau occurrence behind the Earth's orbit. An assumption on a technical hardware error can also be rejected, as the effect was observed by different spacecraft only in the vicinity of the solar minimum, whereas in the maximum, *Ulysses* showed the behavior of the solar wind speed meeting with expectations. Furthermore, if it was a technical effect, an observed shift to higher or lower values would be systematic, rather than quasi-sporadic.

Therefore, the decrease of $V$ in Figs. 3 and 6 can be caused by the existence of a local plasma barrier. It was Parker who was the first to discuss the possibility of formation of barriers in the solar wind [1]. In particular, such a barrier can be formed by the solar wind from coronal holes near the minimum of solar activity. This idea appeared more than 30 years ago [33], but it still remains actual, as it explains the existence of the cavity around the Sun filled with energetic particles. If to accept the hypothesis on the inclination of flows from coronal holes from both poles of the Sun to the helioequator in the form of Russian hills or a fountain and mixing the flows at larger distances, then the observed deceleration of the solar wind at some distances looks quite natural.

Figure 9 illustrates the occurrence of three main zones in the solar wind, following Fig. 21 from [33]. Fast coronal hole flows incline to low latitudes and interact with the slow solar wind, forming so-called co-rotating interaction regions at low heliolatitudes. Closer to the Sun, they expand rather freely in the regular stream zone but finally merge at a certain distance. It was previously thought that this merging occurred at about 8 AU; however, modern 3D MHD models of the solar wind (like ENLIL) that use a real situation on the Sun as boundary conditions predict the stream merging much closer to the Sun, not farther than 5.5 AU [34]. Examples of such behavior of streams from coronal holes can be found at https://iswa.gsfc.nasa.gov/IswaSystemWebApp/ in any phase of the solar cycle, not necessarily in the solar minimum. In the area of merging, a plasma barrier is formed that gradually turns with distance into the area where turbulence and small-scale processes, such as wave interaction, dominate. In general, this scenario is consistent with theoretical predictions on the evolution of turbulence in the solar wind, as the best correspondence to observations is found to occur only beyond 10 AU [3].

Such an approach to the explanation of the effect of the solar wind deceleration at several AU is phenomenological and does not demand a development of theories. However, another approach based on the detailed consideration of features of streams coming to low heliolatitudes is possible. In this work, we discuss a possibility of the impact of spatial irregularities of the solar wind in the corona, including the polar corona, on the solar wind speed profile in the ecliptic plane. We have shown that a local minimum of the solar wind speed observed beyond the Earth's orbit can be determined by the impact of magnetic field lines that left areas with the lowered speed near the borders of coronal holes and reached low heliolatitudes far from the Sun. The model we built to explain the effect is quite simple and demands a further development; however, it reflects the main features of observations, including the difference between solar wind properties in the northern and southern heliosphere.

The effect also can be determined by the potential impact of dust in the asteroid belt located between Mars and Jupiter on the propagating solar wind. We only note such a possibility, not developing a theory, as a separate study is needed to check this hypothesis, which is beyond the scope of this article. An estimation of the solar wind deceleration due to its interaction with dust and larger fragments of the asteroid belt will be performed in the future.

In addition to the main effect discussed above, we found many interesting features of the solar wind speed evolution with distance, heliolatitude, and solar cycle. All the results can be summarized as follows.

1. Numerous deviations of the solar wind speed radial evolution from its expected smooth growth and reaching a plateau beyond the Earth's orbit are found.

2. At low heliolatitudes, within the belt of $\pm40°$, the profile of the solar wind speed resembles Parker's $V(r)$ dependence only in the maximum of the solar activity. The solar wind speed is almost independent of the heliolatitude $\theta$ in the solar maximum at low heliolatitudes.

3. At low heliolatitudes in the minimum of solar activity, the solar wind speed profile shows a decrease between the Earth and Jupiter. The tendency to the deceleration of the solar wind at low heliolatitudes in the minimum of solar activity is confirmed by independent measurements from other spacecraft. The solar wind speed is independent of heliolatitude only within a narrow belt of $\pm10°$ around the helioequator, above which $V$ smoothly grows. There is no essential difference between the $V(\theta)$ profiles in the southern and northern hemispheres during the minimum of solar activity at low heliolatitudes.

4. In the high-latitude heliosphere in the solar activity maximum, there are two steady branches of the solar wind dependence on distance: the falling branch and the growing one. They are determined by a prominent asymmetry of the $V(\theta)$ profiles in the northern and southern hemispheres. The solar wind speed $V$ grows in the northern hemisphere toward the pole, while it decreases toward the pole in the solar maximum in the southern hemisphere.

5. In the high-latitude heliosphere in the minimum of solar activity, there is only one branch of the radial dependence of the solar wind speed, which corresponds to the well-known behavior of $V$ inside coronal holes, i.e., there is a small $V$ growth toward the coronal hole center. The two branches of the $V(\theta)$ profile practically merge in the minimum of solar activity at high heliolatitudes; however, $V$ still grows in the southern hemisphere more slowly than in the northern hemi-sphere as approaching the pole, similar to its behavior in the solar maximum.

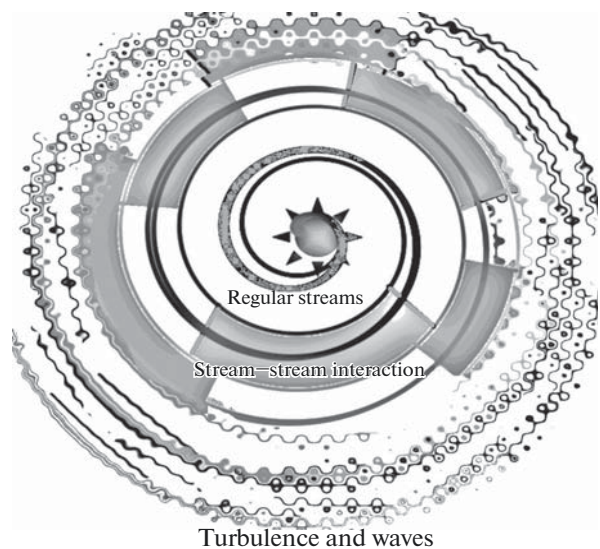

**Fig. 9.** Sketch illustrating the formation of separated zones in the inner heliosphere near the ecliptic plane: zone of free propagation of streams and flows, zone of stream–stream interaction and formation of compression waves, and zone of wave interaction where turbulence and small-scale processes dominate. The idea is adapted from [33], Fig. 21.

6. It is shown that the problem of the explanation of the nonmonotonic radial evolution of the solar wind speed can be solved through the study of spatial characteristics of the solar wind in the solar corona.

## ACKNOWLEDGMENTS

*Ulysses* data were obtained from the official website of the Goddard Space Flight Center OMNIweb plus: http://omniweb.gsfc.nasa.gov. For reconstructions of the solar wind speed shown in Fig. 7a, coronographic observations from *SOHO/UVCS*, and *LASCO* were used (https://sohowww.nascom.nasa.gov/data/data.html), and the Mauna Loa ground observatory data contributed as well (https://www.esrl.noaa.gov/gmd/obop/mlo/). Figure 7b is based on WSO magnetograms (http://wso.stanford.edu).

This work was performed with a financial support from the Russian Science Foundation, grant no. 14-12-00824. It is also a part of the ISSI International Team 405 project "Current Sheets, Turbulence, Structures and Particle Acceleration in the Heliosphere." O.V. Khabarova was supported by RFBR grant nos. 16-02-00479, 17-02-01328, and partially 17-02-00300. V.N. Obridko, R.A. Kislov, H.V. Malova, and V.D. Kuznetsov were supported by RFBR grant no. 17-02-01328. V.N. Obridko additionally thanks RFBR grant no. 17-02-00300. L.M. Zelenyi was supported by RFBR grant no. 16-02-00479.

H.V. Malova was partially supported by RFBR grant no. 16-52-16009-NTsNIL-a and by programs 28 and I.24P of the Presidium of the Russian Academy of Sciences.